\documentclass[aps,prb,twocolumn]{revtex4}
\usepackage{graphicx}

\begin{document}

\title{Anomalous Paramagnetic Effects in the Mixed State of LuNi$_{2}$B$_{2}$C}
\author{Tuson Park}
\author{A. Malinowski}
\author{M. F. Hundley}
\author{J. D. Thompson}
\affiliation{Los Alamos National Laboratory, Los Alamos, New Mexico 87545, USA}
\author{Y. Sun}
\altaffiliation{Current address: Institute of Physics Chinese Academy of Sciences, Beijing 100080, People's Republic of China}
\author{M. B. Salamon}
\affiliation{Department of Physics and Materials Research Laboratory, University of
Illinois at Urbana--Champaign, Urbana, Illinois 61801, USA}
\author{Eun Mi Choi}
\author{Heon Jung Kim}
\author{Sung-Ik Lee}
\affiliation{National Creative Research Initiative Center for Superconductivity and
Department of Physics, Pohang University of Science and Technology, Pohang
790-784, Republic of Korea}
\author{P. C. Canfield}
\author{V. G. Kogan}
\affiliation{Ames Laboratory, Department of Physics and Astronomy, Iowa State University, Ames, Iowa 50011, USA}
\date{\today}

\begin{abstract}
Anomalous paramagnetic effects in \textit{dc} magnetization were observed in
the mixed state of LuNi$_{2}$B$_{2}$C, unlike any reported previously. It
appears as a kink-like feature for $H\geq 30$~kOe and becomes more prominent
with increasing field. A specific heat anomaly at the corresponding temperature
suggests that the magnetization anomaly is due to a true bulk transition. A magnetic flux transition from a square to an hexagonal lattice is consistent with the
anomaly.
\end{abstract}

\maketitle

In cuprate (high-$T_{c}$) superconductors, high-transition temperatures ($T_{c}$) and short coherence lengths ($\xi $) lead to large thermal
fluctuation effects, opening a possibility for melting of the flux line
lattice (FLL) at temperatures well below the superconducting transition
temperature. A discontinuous step in \textit{dc} magnetization and a sudden,
kink-like drop in resistivity signified the first order nature of the melting
transition from the vortex lattice into a liquid.\cite{pastoriza94,zeldov95,welp96} In conventional type~II superconductors, with modest transition temperatures and large coherence lengths, vortex melting is also expected to occur in a very limited part of the phase diagram,\cite{eilenberger67} but it has yet to be observed experimentally. In the rare-earth nickel borocarbides RNi$_{2}$B$_{2}$C (R = Y, Dy, Ho, Er, Tm, Lu), the coherence lengths ($\xi \cong 10^{2}\mathring{A}$) and superconducting transition temperatures (16.1~K for R = Lu) lie between these extremes, suggesting that the vortex melting will be observable and may provide further information on vortex dynamics. Indeed, Mun et al. \cite{mun96} reported the observation of vortex melting in YNi$_{2}$B$_{2}$C, based on a sharp, kink-like drop in electrical resistivity.

Recently, a magnetic field-driven FLL transition has been observed in the
tetragonal borocarbides.\cite{yaron96,eskildsen98,sakata00,eskildsen2001}
The transition from square to hexagonal vortex lattice occurs due to the
competition between sources of anisotropy and vortex-vortex interactions. The
repulsive nature of the vortex interaction favors the hexagonal Abrikosov
lattice, whose vortex spacing is larger than that of a square lattice. The
competing anisotropy, which favors a square lattice, can be due to lattice
effects (fourfold Fermi surface anisotropy), \cite{kogan97} unconventional
superconducting order parameter, \cite{gilardi02} or an interplay of the
two.\cite{nakai02,tuson04} In combination with non-negligible fluctuation
effects, the competition leads to unique vortex dynamics right below the $%
H_{c2}$ line in the borocarbides, namely a reentrant vortex lattice
transition.\cite{eskildsen2001} Fluctuation effects near the upper critical
field line wash out the anisotropy effect, stabilizing the Abrikosov
hexagonal lattice.\cite{gurevich01,klironomos03} Here, we report
the first observation of paramagnetic effects in the \textit{dc}
magnetization $M$ of the mixed state of LuNi$_{2}$B$_{2}$C. The kink-like
feature in $M$ and the corresponding specific heat feature for $H\geq 30$~kOe
signify the reentrant FLL transition, which is consistent with the low-field
FLL transition line inferred from small angle neutron scattering (SANS).\cite{eskildsen2001}

Single crystals of LuNi$_{2}$B$_{2}$C were grown in a Ni$_{2}$B flux as
described elsewhere \cite{cho95} and were post-growth annealed at $T=1000$ C$%
^{\circ }$ for 100 hours under high vacuum, typically low $10^{-6}$ Torr.
\cite{miao02} Samples subjected to a preparation process such as grinding,
were annealed again at the same condition as the post-growth annealing. A
Quantum Design magnetic property measurement system (MPMS) was used to
measure \textit{ac} and \textit{dc} magnetization while the heat capacity
option of a Quantum Design physical property measurement system (PPMS) was
used for specific heat measurements. Electrical resistivity was measured by
using a Linear Research \textit{ac} resistance bridge (LR-700) in
combination with a PPMS.

The detailed \textit{dc} magnetization of LuNi$_{2}$B$_{2}$C reveals an
anomalous paramagnetic effect for $H\geq $30~kOe, where the magnetic
response deviates from a monotonic decrease and starts to rise, showing
decreased diamagnetic response. The in-phase and out-of-phase components of
the \textit{ac} susceptibility $\chi _{ac}$ show a dip and the specific heat
data show a feature at the corresponding temperature, reminiscent of vortex
melting in high-$T_{c}$ cuprates.\cite{schilling97} Electrical transport
measurements, however, do not exhibit any feature corresponding to the
paramagnetic effect; \textit{e.g.}, a sharp drop in the electrical
resistivity. The zero-resistance transition, rather, occurs at a much higher
temperature, suggesting that the paramagnetic effect is not related to
vortex melting. It is instead consistent with a topological FLL change
between square and hexagonal structures.

\begin{figure}[tbp]
\centering  \includegraphics[width=8.5cm,clip]{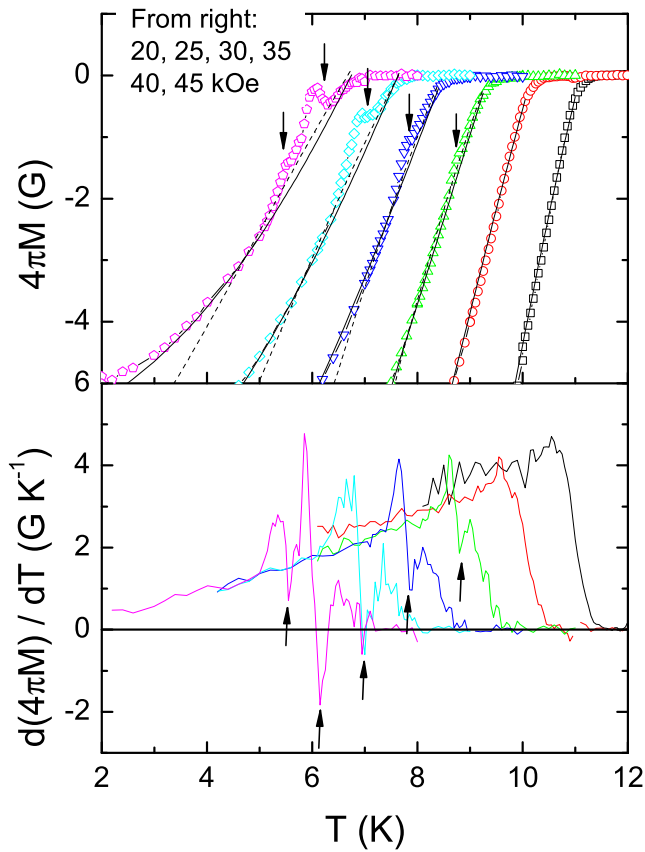}
\caption{(Color online). Top panel: \textit{dc} magnetization $M(T)$ as a function of
temperature at 20, 25, 30, 35, 40, and 45~kOe. Dashed lines are fits from
the standard local London model \protect\cite{degennes66} and solid lines from
the non-local London model \protect\cite{gurevich01} (see text).Bottom
panel: temperature derivation of $M(T)$ at corresponding magnetic fields.
Arrows indicate the points where kink-like features start to appear}
\label{figure1}
\end{figure}
The top panel of Fig.~1 shows \textit{dc} magnetization $M$ as a function of
temperature at several fields. For $H\geq 30$~kOe, kink-like features
appear, which are marked by arrows. The anomalous increase can be easily
seen as a sharp drop in $dM/dT$ (arrows in the bottom panel). The
magnetization reported here is independent of time and has no hysteresis
between zero-field cool (ZFC) and field cool (FC) data within experimental
accuracy, indicating that the measured value is an equilibrium
magnetization. In the top panel of Fig.~1, we compare the data and some
model calculations. Dashed lines are predictions from the standard local
London model: \cite{degennes66} 
\begin{equation}
-4\pi M=M_{0}\ln (\eta H_{c2}/H),
\end{equation}%
where $M_{0}=\phi _{0}/8\pi \lambda ^{2}$, $\eta $ is a constant of order
unity, $\lambda $ the penetration depth, and $\phi _{0}$ the flux quantum.
In the fit, $H_{c2}$ was determined from our resistivity data (see Fig.~2)
and $M_{0}$ from $H_{c2}$ with $\kappa =\lambda /\xi =15$. In order to get the
best result, the fitting parameter $\eta $ was varied between 0.95 and 0.97
and the absolute amplitude of $M_{0}$ was changed as a function of
magnetic field. The local London model explains the monotonic decrease with
decreasing temperature, but the fit becomes worse at higher field. In a
clean system like LuNi$_{2}$B$_{2}$C where the electronic mean free path is
long compared to the coherence length $\xi _{0}$, the current at a point
depends on magnetic fields within a characteristic length $\rho $, or
nonlocal radius. Taking into account the nonlocal current-field relation in
superconductors, a non-local London model was suggested: \cite{kogan96} 
\begin{equation}
-4\pi M=M_{0}[\ln (1+H_{0}/H)-H_{0}/(H_{0}+H)+\Lambda ],
\end{equation}%
where $H_{0}=\phi _{0}/(4\pi ^{2}\rho ^{2})$ and $\Lambda =\eta _{1}-\ln
(1+H_{0}/\eta _{2}H_{c2})$ with $\eta _{1}$ and $\eta _{2}$ being order of
unity. It is worth noting that the scaling parameter $H_{c2}$ in the local
theory is replaced by $H_{0}$ in the non-local model. The nonlocal radius $%
\rho $ slowly decreases with increasing temperature and is suppressed
strongly by scattering. The solid lines are best results from the model
calculation where we used the temperature dependence of $H_{0}$ and $\Lambda 
$ from the literature for YNi$_{2}$B$_{2}$C.\cite{song99} Both the local and
the non-local models explain the temperature dependence of $M$ at low fields,
while only the non-local model can describe the data at and above 35~kOe.
The good fit from the nonlocal model is consistent with the equilibrium
magnetization analysis of YNi$_{2}$B$_{2}$C, \cite{song99} suggesting the
importance of nonlocal effects in the magnetization. The many fitting
parameters in both fits, however, prevent us from making a definite
conclusion as to which model better describes $M(T)$. Nevertheless, we can
extract the important conclusion that the kink-like feature in the mixed
state is a new phenomenon that needs further explanation.

In the early stage of high-$T_{c}$ cuprate research, anomalous paramagnetic
effects in $M(T)$ were reported in the irreversible region and this effect
was later attributed to the field inhomogeneity of the measured scan length
in a SQUID magnetometer.\cite{schilling92} We tested various scan lengths
from 1.8~cm to 6~cm for which the field inhomogeneity varies from 0.005~\% to
1.4~\% along the scan length and found negligible dependence on the measuring
length, which suggests that field inhomogeneity is not the source of the
anomaly. A more definitive test used a conventional type~II superconductor
NbSe$_{2}$ in a similar configuration. There was no such anomalies in NbSe$%
_{2}$ as in the borocarbide. Taken together, we conclude that the reversible
paramagnetic effects are intrinsic to LuNi$_{2}$B$_{2}$C. We also emphasize that the the phenomena is different from the paramagnetic Meissner
effect (PME) or Wohlleben effect \cite{braunisch93} where the FC $\chi $
becomes positive whereas the ZFC $\chi $ remains negative. The PME is an
irreversible effect and occurs in the Meissner state, while the subject of
this study is a reversible effect and takes place in the mixed state.

\begin{figure}[tbp]
\centering  \includegraphics[width=8.5cm,clip]{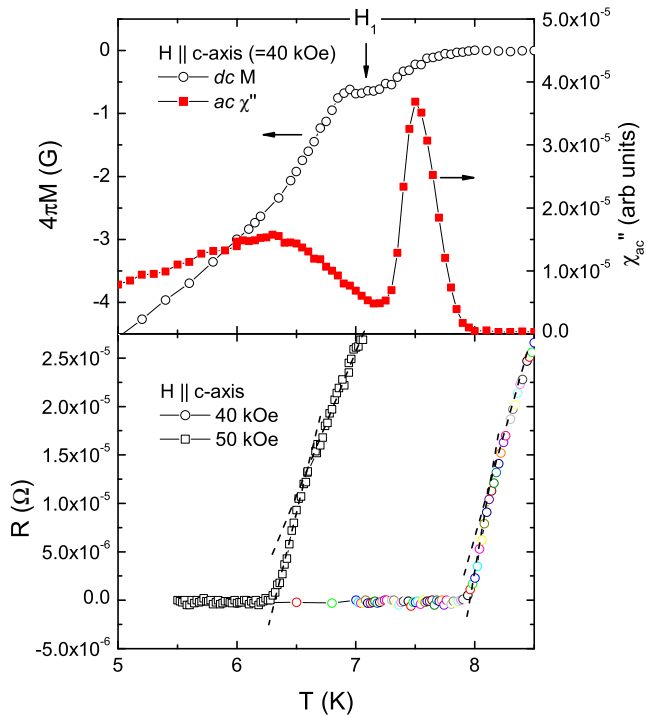}
\caption{(Color online). Top panel: Magnetization (left axis) and the imaginary part of the 
\textit{ac} susceptibility (right axis) at 40~kOe. Bottom panel: Resistive
superconducting transition at 40~kOe (circles) and 50~kOe (squares). The
lines are guide to eyes.}
\label{figure2}
\end{figure}
In the top panel of Fig.~2 the reversible magnetization $M$ (left axis) and
the out-of-phase component of \textit{ac} susceptibility $\chi _{ac}^{\prime
\prime }$ (right axis) are shown as a function of temperature at 40~kOe. A
dip appears both in $\chi _{ac}^{\prime }$ (not shown) and in $\chi
_{ac}^{\prime \prime }$ at the same temperature where $M$ shows the
paramagnetic anomaly. Since a dip in $\chi _{ac}$ is often related to vortex
melting, it is natural to consider the vortex phase change from liquid to
lattice or glass as a possible explanation. The resistive superconducting
transition at 40~kOe (circles) and 50~kOe (squares) are shown in the bottom
panel of Fig.~2. A resistive slope change in the transition region, that can
be considered as a signature of the vortex melting, \cite{mun96} was
observed at 8.1~K and 6.6~K for 40~kOe and 50~kOe, respectively. The $R=0$
transition temperature, however, is much higher than the temperature where
the dip occurs in $\chi _{ac}$, which argues against the vortex melting
scenario as the physical origin of the anomalous paramagnetic effects. The
increase in $M$ at the transition temperature is also opposite from the
decrease in the vortex melting interpretation. 

\begin{figure}[tbp]
\centering  \includegraphics[width=8.5cm,clip]{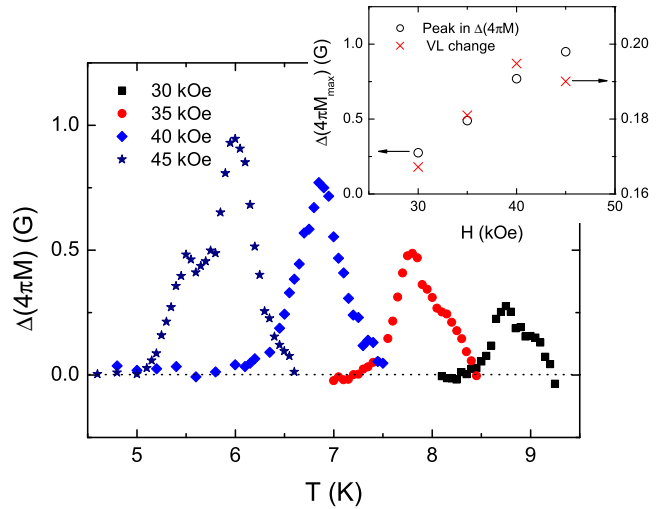}
\caption{(Color online). Temperature dependence of $\Delta (4\pi M)$ at 30, 35, 40, and 45~kOe, where $\Delta M=M-M(nonlocal)$. Inset: The peak values of $\Delta (4\pi M)$ are compared with those estimated from Eq.~(3).}
\label{figure3}
\end{figure}
Recently, a structural phase transition in the FLL was suggested to explain another peak effect observed below the vortex melting line in YBCO. \cite{deligiannis97,rosenstein99} The vanishing of a \textit{squash} elastic mode gives rise to a topological FLL transition and leads to the new peak effect, while the softening of the shear modes $c_{66}$ is relevant to the conventional peak effect in high-$T_{c}$ cuprates.\cite{larkin95} The observation of the dip effect well below the melting line in LuNi$_{2}$B$_{2}$C indicates that the anomalous paramagnetic effects are related to a change in the FLL and the increase in $M$ is also consistent with the FLL change where the Abrikosov geometrical factor $\beta$ changes.\cite{abrikosov57,degennes66} Fig.~3 shows the temperature dependence of the paramagnetic anomaly, $\Delta (4\pi M)$, at several magnetic fields, where $\Delta (4\pi M)$ is the magnetization after subtracting the monotonic, diamagnetic background obtained from Eq.~(2). With increasing field, the peak becomes enhanced and an additional peak is observed at 45~kOe. In extreme type~II material $(\kappa >> 1/\sqrt{2})$, the magnetization change due to a FLL transition is written as
\begin{equation}
\Delta(4\pi M)=\frac{H_{c2}-H}{2\kappa ^{2}-1} \left(\frac{1}{\beta_{\triangle}}-\frac{1}{\beta_{\diamond}} \right),
\end{equation}
where $\beta _{\triangle} \approx 1.16$ for a hexagonal FLL and $\beta _{\diamond} \approx 1.18$ for a square FLL.\cite{degennes66} In the inset of Fig.~3, we compared the peak intensity of $\Delta (4\pi M)$ (left axis) and the estimation from Eq.~3 (right axis) with $\kappa = 15$. It is encouraging to see that the simple model qualitatively reproduces the field dependence of the paramagnetic contribution. However, the quantitative difference in absolute values suggests that a more elaborate model is required.   

\begin{figure}[tbp]
\centering  \includegraphics[width=8.5cm,clip]{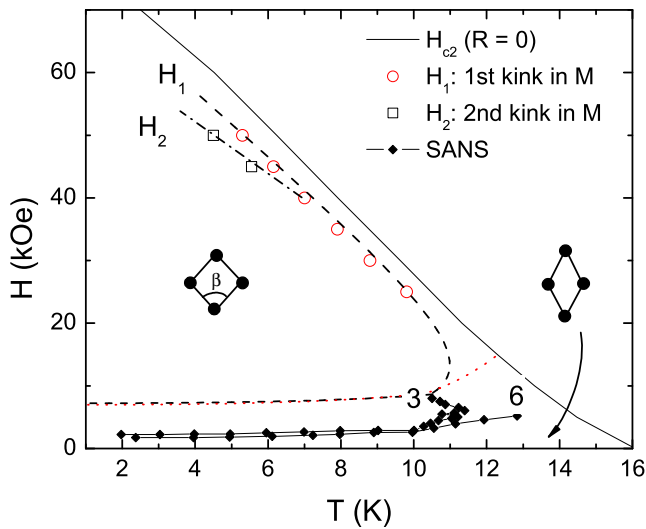}
\caption{(Color online). $H-T$ phase diagram. The $H_{1}$ line indicates the temperature where $M(T)$ deviates from a monotonic decrease and the $H_{2}$ line is where the second anomaly occurs above 45~kOe. The dashed line depicts qualitatively what fluctuation models based on the nonlocal London \protect\cite{gurevich01} or extended GL \protect\cite{klironomos03} predict when the nonlocal radius is approximately $2-3 ~\xi_{0}$. The dotted line is from theories without fluctuation
effects.\protect\cite{affleck97,wilde97} The diamonds are SANS data, where the numbers next to the data indicate the degrees of the azimuthal splitting with which the transition line is determined.\protect\cite{eskildsen2001} The square and rhombic shapes are forms of vortex lattices.}
\label{figure4}
\end{figure}  
The $H-T$ phase diagram is shown in Fig.~4. The upper critical field line $%
H_{c2}$ was determined from the $R=0$ superconducting transition and is
consistent with the temperature where $\chi _{ac}$ starts to have a non-zero
value. The $H_{1}$ line in the mixed state is the point where the reversible
magnetization shows the paramagnetic effects and $H_{2}$ is the 2nd anomaly
that appears above 45~kOe (see Fig.~1). According to the Gurevich-Kogan non-local London model, \cite{gurevich01} the anisotropic nonlocal potential, which is responsible for the low-field FLL transition observed in SANS, \cite{eskildsen2001} is averaged out by thermal vortex fluctuations
near $H_{c2}$. Since the interaction becomes isotropic, the hexagonal
Abrikosov lattice is preferable, leading to the second FLL transition from
square back to rhombic (triangular) lattice as the field gets closer to the $H_{c2}$ line. The reentrant transition is predicted to occur well below the vortex
melting line because the amplitude of vortex fluctuations required to wash
out the nonlocal effects is much smaller than that for the vortex melting.
This prediction is consistent with our observation that the $H_{1}$ line is
much below the $R=0$ transition line. The dashed line depicts
qualitatively what the fluctuations models based on the nonlocal London \cite{gurevich01} or extended GL \cite{klironomos03} predict, which nicely explains the $H_{1}$ line. The dotted line is the FLL transition line that meets the $H_{c2}$ line both in the non-local London model \cite{affleck97} and in the extended Ginzburg-Landau (GL) theory \cite{wilde97} without fluctuation effects. We note that a direct comparison between the SANS and our data is difficult even though they are qualitatively similar. Since $H_{1}$ heavily depends on the sample purity, \cite{gammel99} a factor of 2 or more difference in $H_1$ has been easily observed even among pure compounds.\cite{eskildsen97,eskildsen01} Further, the $H_{1}$ line from the SANS also depends on the criteria used for the FLL transition (see Fig.~4).

For vortex melting, where the lattice changes to a liquid, the transition
involves latent heat and the specific heat shows a sharp peak at the
transition temperature.\cite{schilling97} For a structural change in the vortex lattice, the transition is probably of 2nd order because an infinitesimally small change of the angle $\beta$ between adjacent vortex lines changes the symmetry. Based on the paramagnetic jump (Fig.~1) and $dH_{1}/dT <0$ (Fig.~3), Ehrenfest's relation at constant field predicts a suppression of $C/T$ as the FLL changes from a rhombic to a square lattice. Fig.~5 shows the specific heat data of LuNi$_{2}$B$_{2}$C at 45~kOe as a function of temperature. In addition to the superconducting transition between 6.22 and 6.89~K, an anomaly is, indeed, observed at 6.12~K which corresponds to the anomalous paramagnetic effects. Depending on the back ground we choose, however, the anomaly can be considered as either a suppression or a jump.\cite{jump} Similar features at 40 and 50~kOe were also observed at the temperatures corresponding to the paramagnetic effects in $H_{1}$. More sensitive measurements such as ac calorimetry will help in resolving the issue. Finally, we note that we are not able to discern any corresponding feature to the $H_{2}$ line in $C_{p}$ or in $\chi _{ac}$. More work is in progress to understand the second paramagnetic jump in $M$ which appears for $H\geq 45$~kOe.
\begin{figure}[tbp]
\centering  \includegraphics[width=8.5cm,clip]{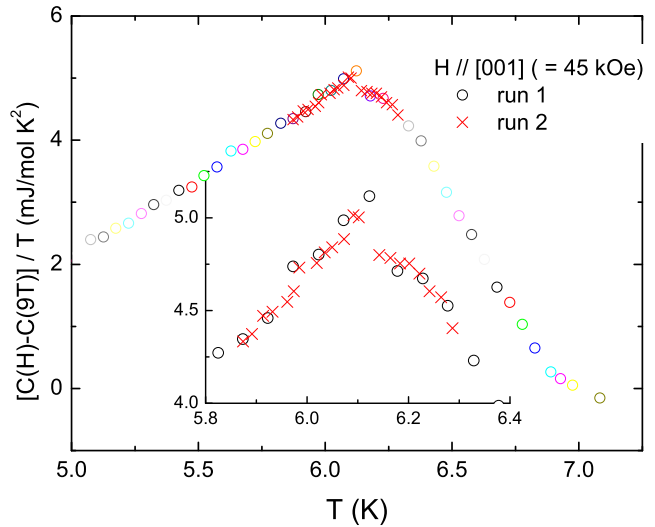}
\caption{(Color online). Specific heat difference $C(H)/T-C(9T)/T$ vs $T$ at 45~kOe for $H\parallel \lbrack 001]$. Inset: blow-up of the main panel at around 6.1~K. Different symbols correspond to different sets of measurements and attest to the reproducibility of these results.}
\label{figure4}
\end{figure}

In summary, we report the first observation of an anomalous paramagnetic
jump in the magnetization of the mixed state of LuNi$_{2}$B$_{2}$C. A dip
appears in $\chi _{ac}$ at the same temperature as the paramagnetic effects,
suggesting the relevance of the flux line lattice. The $H-T$ phase diagram
is consistent with a FLL structural transition from square to hexagonal
lattice just below the upper critical field line. The observation of an
additional feature in the specific heat data at the corresponding temperature underscores the interpretation of paramagnetic effects as due to a reentrant FLL
transition in LuNi$_{2}$B$_{2}$C.

Work at Los Alamos was performed under the auspices of the U.S.
Department of Energy (DOE) and at Urbana under NSF Grant No. DMR 99-72087. The work at Pohang was supported by the Ministry of Science and Technology of Korea through the Creative Research Initiative Program and at Ames by Iowa State University of Science and Technology under DOE Contract No. W-7405-ENG-82. We acknowledge benefits from discussion with Lev N. Bulaevskii, M. P. Maley, and I. Vekhter. We thank T. Darling for assistance in sample annealing.

\bibliography{lnbc}

\begin{thebibliography}{33}
\bibitem[{jum()}]{jump}
\bibinfo{note}{When we extrapolate the specific heat data from above the H1
  transition temperature (= 6.15 K) and consider them as a background, the
  feature at H1 can be considered as a decrease}.
\end{thebibliography}

\end{document}